\documentclass[12pt,reqno]{amsart}

\usepackage{animate}
\usepackage{geometry}                
\usepackage[parfill]{parskip}    
\usepackage{graphicx}
\usepackage{amssymb}
\usepackage{epstopdf}
\usepackage{animate}
\usepackage{tikz}
 \usepackage[colorlinks=true,hyperfootnotes=true,citecolor=cyan]{hyperref}

\usepackage[width=\linewidth,labelfont=small,textfont={it,footnotesize}]{caption}

\usepackage{mathpazo}
\fontfamily{Palatino}

\begin{document}

\newcommand{\be}{\begin{equation}}
\newcommand{\ee}{\end{equation}}
\newcommand{\ba}{\begin{eqnarray}}
\newcommand{\ea}{\end{eqnarray}}

\newcommand{\dd}{{\rm d}}
\newcommand{\rs}{r_{\rm s}}
\newcommand{\as}{a_{\rm s}}
\newcommand{\rl}{r_\star}
\newcommand{\tl}{t_\star}
\newcommand{\rhos}{\rho_\star}
\newcommand{\Lambdae}{\Lambda_{\rm e}}

\newcommand{\mU}{{\mathcal{U}}}
\newcommand{\VB}{{\mathcal{V}}}
\begin{center}
\small{Essay written for the Gravity Research Foundation 2020
Awards for Essays on Gravitation} \\
\vspace{2cm}
\Large{\bf Screening away the $H_0$ tension} \\
\vspace*{1cm}
\large{\bf  Jose Beltr\'an Jim\'enez$^\dagger$, Dario Bettoni$^\dagger$ and Philippe Brax$^\ddag$}  \\
\vspace{1cm}
\end{center}
\small{
$^\dagger$ {\it Departamento de F\'isica Fundamental and IUFFyM, Universidad de Salamanca, E-37008 Salamanca, Spain.}\\
\vspace{0.1cm}
$^\ddag$ {\it Universit\'e Paris-Saclay, CNRS, CEA, Institute de physique th\'eorique, 91191, Gif-sur-Yvette, France.}}
\vspace{0.2cm}

\vspace*{0.5cm} 
{\bf Abstract:} This Essay explores  consequences of a dark non-linear electromagnetic sector in a Universe with a net dark charge for matter. The cosmological dynamics can be described by a Lema\^itre model and understood thanks to a screening mechanism driven by the electromagnetic non-linearities that suppress the dark force on small scales. Only at low redshift, when the screening scale enters the Hubble horizon, do cosmological structures commence to feel the dark repulsion. This repulsive force enhances the local value of the Hubble constant, thus providing a promising scenario for solving the Hubble tension. Remarkably, the dark electromagnetic interaction can have a crucial impact on peculiar velocities, i.e. introducing a bias in their reconstruction methods, and having the potential to explain the presence of a dark flow.
\\

{\bf Date}: March 31st, 2020.

\vspace*{0.5cm}
\noindent
\rule[.1in]{4.5cm}{.002in}

\noindent \href{mailto:jose.beltran@usal.es}{jose.beltran@usal.es} ({\it Corresponding author}) \\
\noindent  \href{mailto:bettoni@usal.es} {bettoni@usal.es}\\
\noindent \href{mailto:philippe.brax@ipht.fr}{philippe.brax@cea.fr}

\newpage

The predominant role played by gravity on the large scale dynamics of the Universe finds its roots in its long-range character. Moreover, gravitation being universally attractive,  all  masses contribute constructively to the gravitational interaction. The other known long-range force, electromagnetism, behaves drastically differently and remains marginal on cosmological scales as charges tend to neutralize in an uncharged Universe. This is the fundamental reason why only gravity matters for the evolution of the Universe. However, the advent of dark energy has led to speculations on the existence of additional long-range interactions that could play an important role in cosmology and potentially account for the cosmic acceleration. A tension arises, however, because no such interactions beyond the four fundamental forces has ever been detected in laboratory, solar system experiments or in astrophysical systems like  binary pulsars or in the emission of GWs. In all cases, agreement with the predictions of GR is non-trivial as these new putative interactions must exhibit screening mechanisms hiding them from local gravity tests. Modern models equipped with such screening mechanisms have been developed for scalar fields\footnote{We employ scalar field in a broad sense referring to e.g. a fundamental scalar field, a scalar mode in higher dimensional scenarios, the helicity-0 mode of some higher spin field...}. All in all, screening mechanisms have a very long history and the first one ever is at the very heart of the non-linear Born-Infeld electromagnetism \cite{Born:1934gh}, built in the early 1930's as a pre-quantum regularization of point-like charges. Later in the 1980's, the interest in this theory was revived as it appeared in the low energy limit of open string theory \cite{FRADKIN1985123}. In Born-Infeld electromagnetism, the electric force is Maxwellian at distances larger than a given scale $\rs$, but below such a scale the force is strongly suppressed. It is now clear that the screening at work in Born-Infeld electromagnetism is also shared by a broad class of non-linear electrodynamics, see Table \ref{Table:examples}. A crucial difference between an interaction mediated by a scalar and electromagnetism is that, while the former is always attractive, the latter leads to a repulsive force for like charges. This will be fully exploited in the following.

\begin{figure}[ht]
 \includegraphics[width=0.49\linewidth]{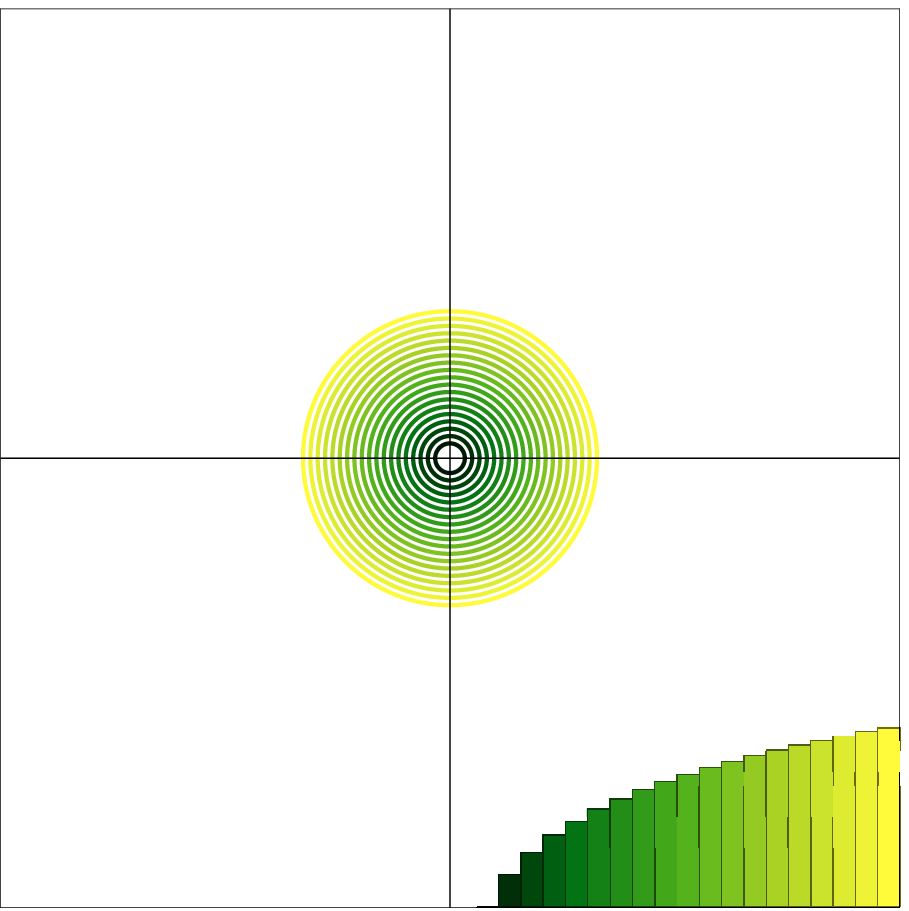}
  \includegraphics[width=0.49\linewidth]{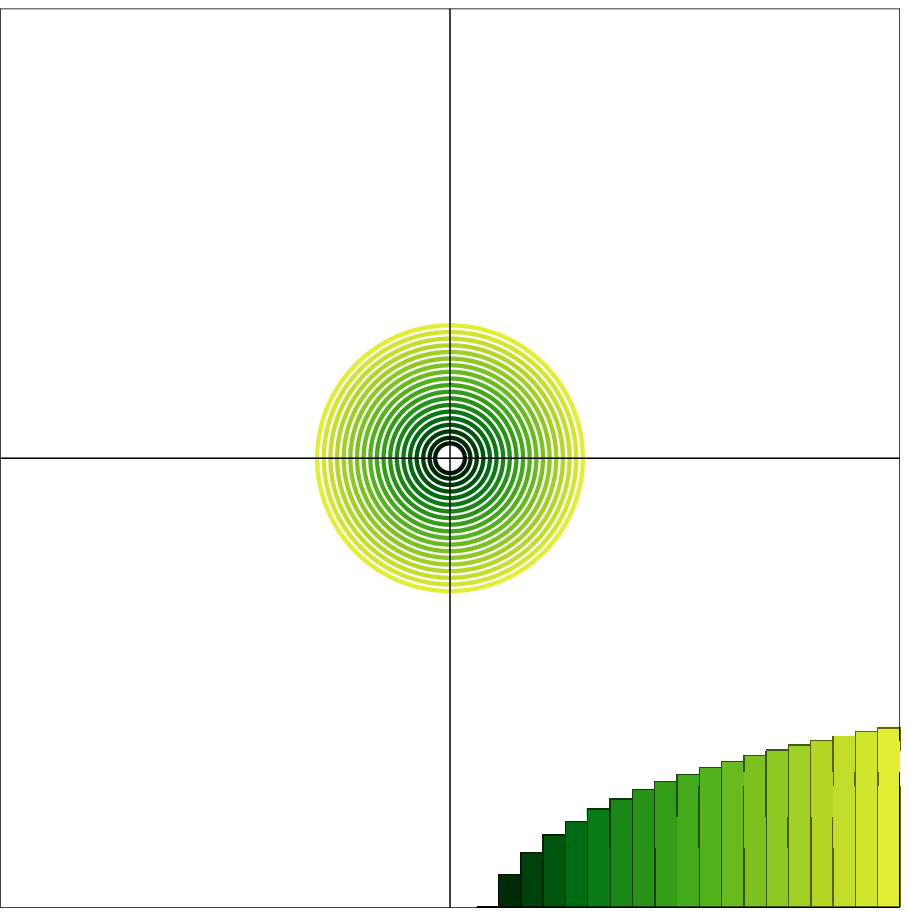}
   \includegraphics[width=0.49\linewidth]{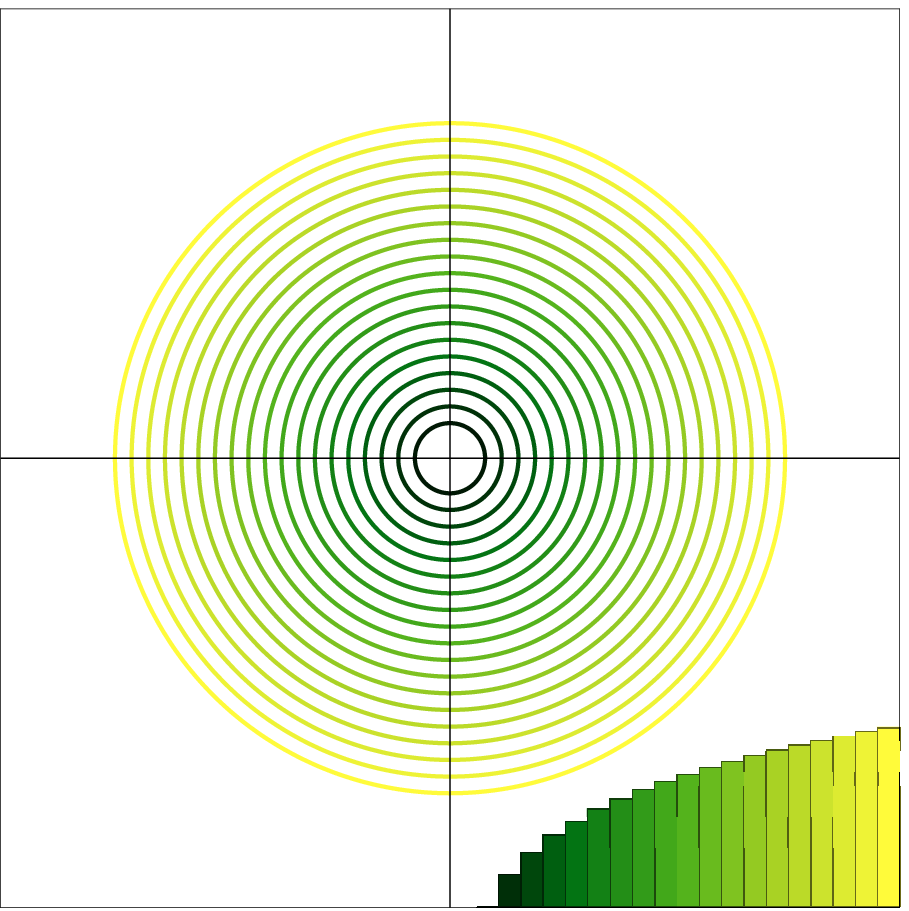}
    \includegraphics[width=0.49\linewidth]{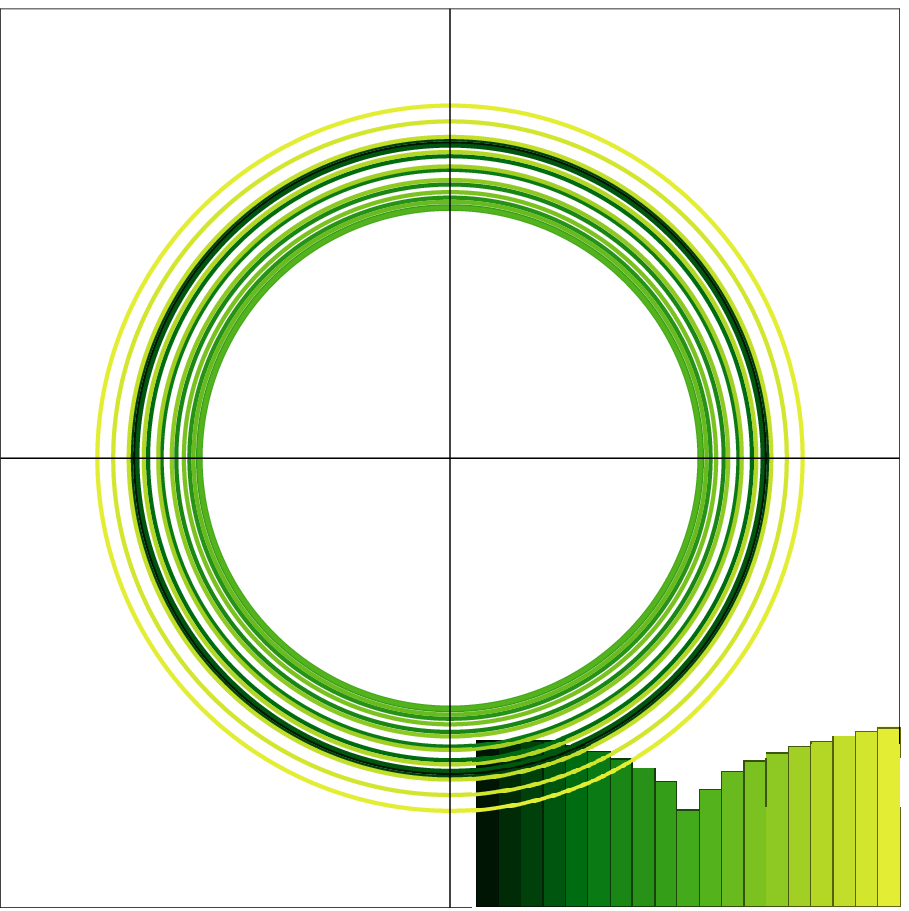}
    \caption{These plots show, in arbitrary units, the numerical solution for the shells without (left) and with (right) shell crossing together with the (log) mass distribution (insets) for the initial time (upper) and some subsequent time (lower). We have used a discretization with 20 shells and assigned an initial mass to each shell consistent with a uniform profile so that $M_i\propto r_i^3(\tl)$. The appearance of shell crossing depends both on the force profile and the interaction strength. Since the inner shells exit their screening scale earlier than the outer shells, the shell crossing is determined by how much faster the inner shells expand due to the extra-repulsion before the outer shells cross their $\rs$. We can corroborate how the mass distribution in the left panels without shell crossing remains constant, while the right panels with shell crossing present a change in the mass distribution as the shells cross. Interestingly, it is the very repulsive nature of the force what can lead to shell crossing in the expanding phase. This cannot happen for models with scalar fields due to their intrinsically attractive character.}\label{Fig:shells}
  \end{figure}

Using a  Newtonian cosmology approach\footnote{See the insightful discussion in \cite{HARRISON1965437} and the rigorous analysis of the discretized approach to Newtonian cosmology in \cite{Gibbons:2013msa}.}, that provides an accurate description of a dust dominated Universe, we can easily include the effects of dark non-linear electromagnetism on the dynamics of the Universe. Let us consider an isotropic initial distribution of equally charged particles with  with a uniform density profile $\rhos$. We will also assume isotropic initial velocities so the spherical symmetry is preserved throughout. The  equation of motion for the spherical shell at position $R$ is\footnote{Spherical symmetry prevents magnetic fields even though there are moving charges, hence no Lorentz force. The Poynting vector remains trivial and there is no electromagnetic radiation.} 
\be
\ddot{R}=-\frac{GM(R)}{R^2}\Big[1-\beta F(R/\rs)\Big]
\ee
with $M(R)$ the mass enclosed within a sphere of radius $R$, $\beta$ the relative strength (per unit mass) of the additional force with respect to gravity and $F(x)$ an interpolating function with $F( x\gg 1)\rightarrow 1$ and $F(x\ll 1)\rightarrow0$. This behavior captures in a phenomenological manner the screened solutions for the dark force without resorting to any specific realization of the non-linear theory. In Table \ref{Table:examples} we provide  explicit examples  (see \cite{Jimenez:2020bgw} for more details).
\begin{center}
\begin{table}
\begin{tabular}{|c|c|c|c|}
 Theory & Lagrangian & Function & Screening scale \\ \hline\hline
   & & &\\
 Born-Infeld &$\frac{\mathcal{L}_{\rm BI}}{\Lambdae^4}=1-\sqrt{-\det\big(\eta_{\mu\nu}+\frac{1}{\Lambdae^2}F_{\mu\nu}\big)}$  & $F(x)=\frac{1}{\sqrt{1+x^{-4}}}$&$\rs=\frac{1}{\Lambdae}\sqrt{\frac{Q}{4\pi}}$ \\&&&\\ 
 &&&\\
 Quadratic & $\mathcal{L}_2 =-\frac{1}{4}F_{\mu\nu} F^{\mu\nu}+\left(\frac{F_{\mu\nu}F^{\mu\nu}}{4\Lambda^4}\right)^2$ & $\left[1+\frac{F(x)}{x^4}\right]F(x)=1$   &$\rs=\frac{1}{\Lambdae}\sqrt{\frac{Q}{4\pi}}$ \\&&&\\\hline
\end{tabular}
\vspace{0.1cm}
 \caption{We give two specific examples of non-linear electromagnetism featuring a screening mechanism. Accidentally, the screening scale coincides for these two theories, but the parametric scaling with $Q$ and $\Lambdae$ is universal. In general, a non-linear theory described by $\mathcal{L}=\mathcal{K}(Y)$ with $Y=-\frac{1}{4}F_{\mu\nu} F^{\mu\nu}$ generates an electric field for a charge $Q$ determined by Gauss' law: $\mathcal{K}_Y \vec{E}=\frac{Q}{4\pi r^3}\vec{r}$. The electric force on a particle of charge $q$ is then $\vec{F}_e=q \vec{E}$. Since we assume all particles to be equally charged, the charge of a given shell is proportional to its mass and the screening radius scales as $\rs=\lambda \sqrt{M}$, with $\lambda$ some constant. All the charges and proportionality constants are absorbed into the parameter $\beta$ in \eqref{eq:app}.}.\label{Table:examples}
\end{table}
\end{center}
Instead of using the Eulerian coordinate $R(t)$, it is convenient to introduce the Lagrangian coordinate $r=R(\tl)$ at  a given  initial time $\tl$ and define the scale factor $a(t,r)\equiv R(t)/r$ so we have
\be\label{eq:app}
\ddot{a}=-\frac{G\mu(t,r)}{a^2}\Big[1-\beta F(a/\as)\Big]
\ee
with $\mu\equiv M r^{-3}$ and $\as=\rs/r$. Since the screening radius scales with the mass as $\rs=\lambda\sqrt{M}$, we have that $\as=\lambda\sqrt{\mu r}$. A central feature for the evolution is the shell crossing condition. In the absence of shell crossing, the mass inside a given shell is conserved and $\mu$ has the constant value
\be
\mu=\mu(\tl)=4\pi \rhos \int_{r< \rl} \frac{r^2\dd r}{\rl^3} =\frac{4\pi \rhos}{3}.
\ee
All the dependence on $r$ is then absorbed by $F$. In this situation, the phenomenology can be easily understood in terms of the screening scale. Well below the screening scale $a\ll\as$, we can neglect the dark interaction and we recover the usual result that the shells evolve in a co-moving motion as in a dust dominated Universe. As the shells start crossing their screening radii $\rs$, self-similarity is broken and the co-moving motion ceases, leading to  an inhomogeneous density profile (see Animation \ref{Fig:shells}).

\begin{figure}[ht]
\vspace{0.2cm}
\includegraphics[width=0.49\linewidth]{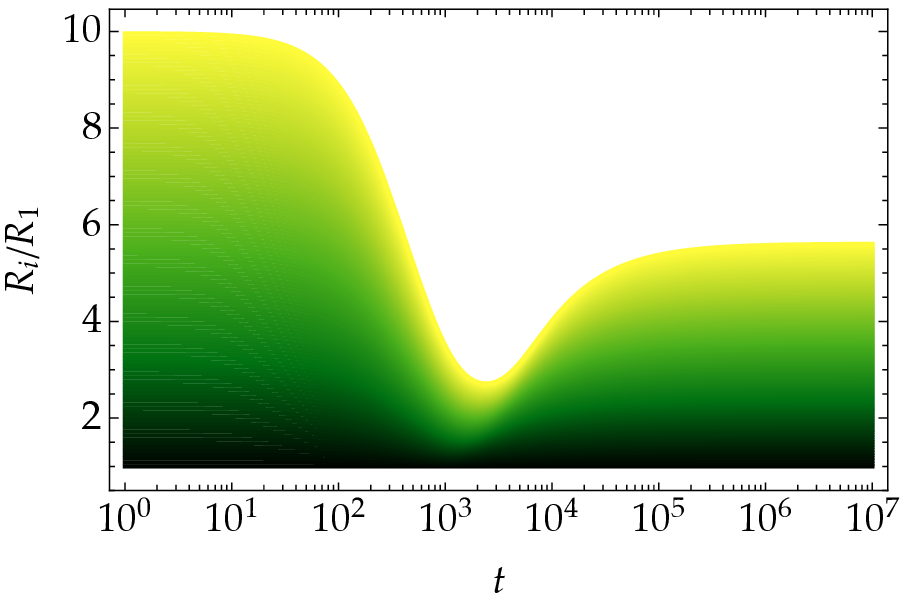}
\includegraphics[width=0.49\linewidth]{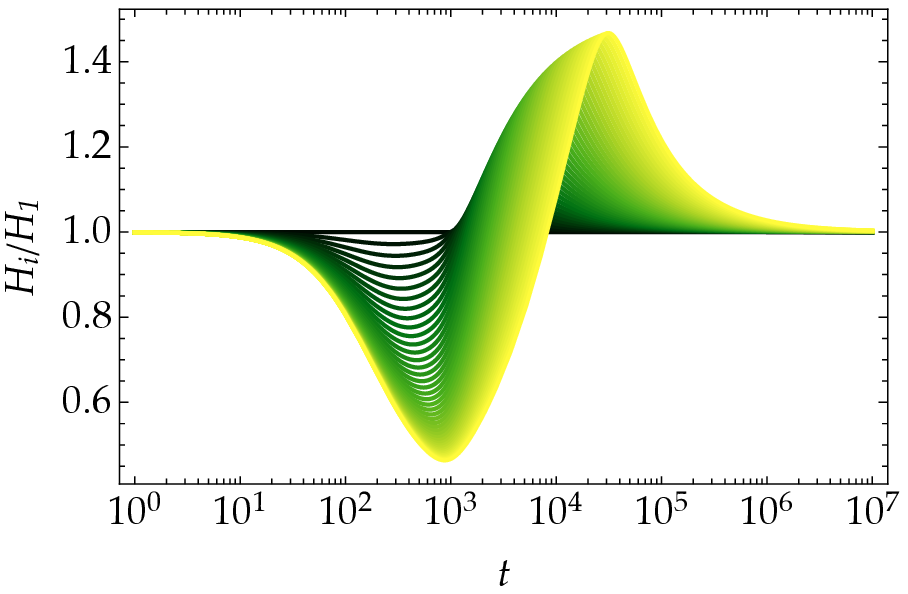}
\caption{In this plot we show the evolution of the radii of the shells (left) and their Hubble expansion rates (right) normalized to their values for the innermost shell. We have used a Born-Infeld model with $\beta=200$, $\rs=500\sqrt{M}$ and an initial profile $4\pi G\rhos/3=1$. We can clearly see how, as the shells start exiting their screening radii, the expansion of the outer shells slows down compared to the one of the inner shells. Similarly the Hubble expansion rate decreases for the larger shells. This suggests how local measurements of the Hubble constant would result in larger values as compared to the cosmological values. Eventually, all the shells exit their screening radii and the expansion rate becomes homogeneous again. See \cite{Jimenez:2020bgw} for a more detailed discussion, including the effect of adding uncharged baryons.}
\label{Fig:H}
\end{figure}

We can find a first integral of motion  corresponding to the energy function given by
\be
E(r)=\frac12\dot{a}^2-\frac{4\pi G\rhos}{3 a}-\mU(a)
\ee
with
\be
\frac{\dd\mU(a)}{\dd a}=\frac{4\pi G\rhos}{3a^2}\beta F(a/\as).
\label{eq:U}
\ee
 We can rewrite the energy equation in the more suggestive form
 \be
 \frac{\dot{a}^2}{a^2}=\frac{8\pi G\rhos}{3 a^3}+2\frac{E(\rl)+\mU(a)}{a^2}.
 \label{Eq:InhomogeneousH}
 \ee
We can then define the inhomogeneous Hubble factor $H(t,r)=\dot{a}/a$ that captures the evolution of the distribution. It is now straightforward to understand how the Universe evolves on different scales. Firstly, let us notice that the constant mode $\mU_0=\mU_0(r)$ determined by boundary/initial conditions simply contributes to the inhomogeneous spatial curvature $k(r)\equiv  E(r)+\mU_0(r)$. The dynamical part can also be analyzed as follows:
\begin{itemize}
    \item On small scales $F(a\ll\as)\simeq 0$ so only the constant mode $\mU_0(\rl)$ contributes. A purely dust Universe with spatial curvature is recovered. On these scales, all the dependence on $r$ drops and we recover the comoving expansion.
    
    \item On scales outside the screening radius $F(a\gg\as)\simeq 1$ and we obtain the equation
    \be
    H^2(t,r)\simeq \big(1-\beta\big)\frac{8\pi G\rhos}{3 a^3}+2\frac{E(r)+\mU_0(r)}{a^2}.
    \ee
    The redressed Newton constant can be equivalently attributed to the additional electrostatic repulsion or to the inclusion of the negative electrostatic binding energy in the total mass. As the crossing times depend on $\rs$, each shell acquires a different spatial curvature, which dominates the overall expansion rate, despite the factor $1-\beta$, making the inner shells expand faster. Eventually, when all shells have exited their screening radii, the expansion rate becomes asymptotically homogeneous, although an inhomogeneous density profile remains (see Figure  \ref{Fig:H}).
\end{itemize}

It is noteworthy that \eqref{Eq:InhomogeneousH} reduces to the Newtonian limit of a relativistic Lema\^itre model \cite{Lemaitre:1933gd}. These models are described by the spherically symmetric line element \cite{Bolejko:2011jc}
\begin{equation}
    \dd s^2 = - e^{A(t,r)}\dd t^2 + e^{B(t,r)}\dd r^2 + R^2(t,r)\dd\Omega^2
\end{equation}
with $A,B$  the gravitational potentials and the radial function and coordinate identified to the Eulerian and Lagrangian coordinates defined above. Einstein's equations read
\be
\dot M_{\rm tot}=-4\pi R^2\dot R p_{\rm tot}(r,t),\quad
 M_{\rm tot}'=4\pi R^2 R' \rho_{\rm tot}(r,t),
\ee
where\footnote{We include a cosmological constant here. Its inclusion in the Newtonian analysis is straightforward.}
\be
 2 G M_{\rm tot} = R + R e^{-A} \dot R^2 - R e^{-B} R'^2 - \frac{1}{3}\Lambda R^3.
 \label{Eq:defMtot}
\ee
These equations admit a physical interpretation in terms of the first law of thermodynamics $\dd M=-p\dd V$ and the definition of the total mass. This mass can be split as $M_{\rm tot}=M+M_e$ with $M$ the gravitational mass and $M_e$ the contribution from the electrostatic energy. Since we are assuming pressure-less matter, we have 
\be 
M_{\rm e}= -4\pi \int \dd t p_{\rm e} R^2 \dot R\,,
\label{secondM}
\ee 
that clearly identifies $M_{\rm e}$ with the electrostatic work.

From \eqref{Eq:defMtot}, using the conservation equations and assuming small pressure gradients we can obtain the following equation:
\begin{equation}\label{eq:lam_Newt}
  \frac{\dot R^2}{R^2}= \frac{2 G M_{\rm total}}{R^3}+\frac{1}{3}\Lambda + 2\frac{ E(r)- \VB}{R^2}
\end{equation}
with
\begin{equation}
    \VB=\int\frac{p'}{ \rho+p}\frac{\dot R}{R'}\dd t\,.
\end{equation}
The resemblance of this equation with \eqref{Eq:InhomogeneousH} is more than suggestive and, in fact, upon the identification
\be 
\frac{\VB}{R^2}=\frac{\mU(a)}{a^2}\quad\Rightarrow\quad
\frac{p_{\rm e}'}{\rho+p}=\frac{4\pi \beta G \rhos rR'}{3a^2(r,t)}F(a/\as)
\label{eqfirst}
\ee
we recognize that they both describe the same cosmology. Furthermore, this  makes apparent how the electric interaction generates a pressure gradient above the screening scale where $F\neq 0$.

One of the pressing questions in cosmology is the increasingly significant tension in the  Hubble constant determination as measured from local objects \cite{Riess:2019cxk} compared to the value inferred from CMB data \cite{Aghanim:2018eyx}. Our scenario provides an appealing mechanism to reconcile both values as the local measurements would be probing the Hubble parameter of the innermost shells, while the inferred value from CMB data would provide the value of the outermost (cosmological) shells. As we have seen, the inner shells experience an expansion with a higher value of $H$ due to the additional repulsion mediated by the dark electric force. Physically, the mechanism can be understood in terms of the electrostatic pressure exerted by the inner shells on the outer shells that strengthens the local expansion rate as compared to the cosmological 
one.

Crucially, the additional interaction only acts in the late time universe when the screening radius is smaller than the Hubble horizon. This prevents any impact on the cosmological evolution at high redshift and only the low redshift evolution is affected. An interesting observable consequence of this scenario is the correction due to the dark force to the velocity field of the matter distribution that will not only be determined by the infall into the gravitational wells, but also contains an extra component in the peculiar velocity caused by the dark electrostatic interaction. Hence, in a region of the Universe whose matter distribution is known and if one can reconstruct the velocity field by independent means, one could test the presence of the additional interaction by simply looking for  the presence of an anomalous peculiar velocity. This effect could have an important impact in the determination of the velocities as, if not properly taken  into account, the electrostatic repulsion\footnote{Since the magnetic force is $\mathcal{O}(v^2)$ and the typical velocities are small, magnetic forces give sub-leading corrections.} would lead to  a biased value for the velocities. Interestingly, there have been claims on the existence of a large scale dark flow that does not seem to be attributable to the matter distribution within the $\Lambda$CDM model \cite{Kashlinsky:2008ut,Atrio-Barandela:2014nda}. In our case, the additional electrostatic repulsion can boost the host clusters and therefore would account for the apparently anomalous flow. Because of the repulsive electrostatic force, one would also expect a clear signature in the abundance of satellite galaxies as well as  modified density profiles and distributions for voids. These effects however would necessitate to resort to proper N-body simulations in order to make definite predictions.

Summarizing, the presence of a dark non-linear electromagnetic interaction could alleviate some of the tensions of the standard cosmological scenario as well as providing distinctive signatures in the matter distribution and the velocity field that can be tested with future surveys.

\bibliographystyle{JHEP}
\bibliography{bibliographyEssay}

\end{document}